\documentclass[aps,prb,twocolumn,superscriptaddress,showpacs]{revtex4}
\usepackage{graphicx}% Include figure files
\usepackage{dcolumn}
\usepackage{hyperref}
\usepackage{slashbox}
\hypersetup{pdfstartview=FitH}
\begin{document}

\title{Phonons in an Inhomogeneous Continuum:
Vibrations of an Embedded Nanoparticle}

\author{Daniel B. Murray}
\affiliation{Department of Physics and Astronomy,
Okanagan University College, 3333 College Way,
Kelowna, British Columbia, Canada V1V 1V7}
\email{dmurray@ouc.bc.ca}

\author{Lucien Saviot}
\affiliation{Laboratoire de Recherche sur la R\'eactivit\'e des Solides,
UMR 5613 CNRS - Universit\'e de Bourgogne\\
9 avenue A. Savary, BP 47870 - 21078 Dijon - France}
\email{lucien.saviot@u-bourgogne.fr}

\date{\today}

\begin{abstract}
The spectrum of acoustic vibrational modes of an inhomogeneous
elastic continuum are analyzed with application to a spherical
nanoparticle embedded in an infinite glass block.  The relationship
of these modes to the discrete vibrational spectrum of a free
sphere is studied.  The vibrational modes of a sphere with a
fixed surface are relevant in some situations.  Comparisons
are also made to calculations of mode frequency and damping
based on complex valued frequency.
\end{abstract}

\maketitle

\section{Introduction}
In 1882, Lamb\cite{Lamb1882} calculated the frequencies of
vibration of a free isotropic homogeneous continuous sphere.
Twenty years ago,\cite{Gersten80,Duval86} this model was applied
to nearly spherical atomic clusters a few nanometers in size
(nanoparticles), whose vibrations can be observed by low
frequency inelastic light scattering (Raman or Brillouin\cite{Bassi01,Kuok03}
setups depending on the actual particle size) and also by
femtosecond pump-probe experiments.\cite{voisinPB02}
However, with exceptions,\cite{Kuok03,Gotic96,Ivanda99,Ristic02,Dieguez01}
the free sphere model hardly seems applicable because the
samples investigated are doped glasses, \textit{i.e.}
nanoparticles solidly embedded within a glass matrix.

Variations on Lamb's calculation attempt to
take into account the mechanical properties of the matrix.
The nanoparticle surface can be rigidly fixed.\cite{saviot96}
Assuming outgoing travelling waves in the matrix,
complex valued mode frequencies can be found.\cite{dubrovskiy81,saviotPRB03}
These approaches give discrete sets of vibrational modes,
unlike a macroscopic object in which there is a continuum.

Conversely, the nanoparticle-matrix interface influence
on vibrations can be ignored\cite{Sirenko98} and bulk phonon modes
can be used as if the nanoparticle surface were acoustically
invisible.  This latter approach can be workable when other mechanisms,
such as size quantization of nanoparticle electron energy levels
(excitons), provide the dominant contribution to the qualitative
features of light scattering spectra.

None of the foregoing provides a satisfactory
description of a nanoparticle-matrix system even if they can provide
valuable insight.  What is needed is a calculation that properly
models the effect of the presence of the nanoparticle on the
phonon modes.  In this paper, we present such an approach based
on recent calculations on core-shell systems.\cite{PortalesPRB02}

Apart from some general observations made in section \ref{GIS},
the nanoparticle is idealized as an isolated spherical homogeneous
isotropic classical continuum object embedded in similarly
idealized glass.  Departures from this won't be addressed in this
paper even though they can be experimentally important. Taking
into account nanoparticle size distribution is straightforward.
Anisotropy has been discussed in a previous paper,\cite{saviotPRB03}
and here we use directionally averaged sound speeds
calculated as described therein.  Non-sphericity
is a more complex problem but arbitrary shapes can be investigated
using molecular dynamics simulation.\cite{saviotPRB03}
Application of continuum elastic boundary conditions
at the interface between nanoparticle and matrix is also an
idealization that needs to be justified with attention to
details at the microscopic level. For example, such conditions are
generally incorrect for epitaxial interfaces.\cite{bettenhausen03}

Mode frequencies predicted for a free sphere correspond to
dampened motions at the same frequency of the nanoparticle in a
sufficiently soft and light matrix.  Some qualitatively new modes
appear corresponding to semi-rigid libration (angular oscillation)
and rattling (translational oscillation) of the nanoparticle as a
result of the elastic restoring force provided by the matrix. Mode
frequencies for a sphere with a fixed boundary correspond to the
dampened motions of a nanoparticle in a harder and denser matrix.

\section{General Inhomogeneous System}
\label{GIS}
Although subsequent sections of the paper deal with isotropic
and homogeneous materials, it is useful to first
consider a continuous elastic medium which may be inhomogeneous
and anisotropic.
$\vec{r}(\vec{R},t)$ denotes real space coordinates
of the point of matter which has material coordinate $\vec{R}$.
Deviation from equilibrium is described in terms of the displacement
field $\vec{u}(\vec{R},t) = \vec{r}(\vec{R},t) - \vec{R}$.
The mass density of the medium is $\rho(\vec{R})$.  Note that
this density refers to material coordinates.  The density in real
space coordinates varies as acoustic vibrations in the medium cause
local expansion and contraction.

The elastic constants of the material are $c_{ijkl}(\vec{R})$.
Stress $\sigma_{ij}$ is related to the displacement field through
$\sigma_{ij} = c_{ijkl} u_{k,l}$
where a summation is implied over repeated indices and the comma
implies a partial derivative.  The equation of motion
is $\rho(\vec{R}) \ddot{u_{i}} = \sigma_{ij,j}$.

Six different models, with abbreviations listed
in Table~\ref{tab:abbr},
have been studied.  For some of them (BSM,FSM and CSM)
$\vec{u}$ is real valued.  For the others,
$\vec{u}$ is complex valued, but the imaginary part
of $\vec{u}$ is to be ignored.

The equation of motion has normal modes $\vec{u}_n(\vec{R})$ with
frequency $\omega_n$ where $n$ is a mode index.
The angular frequency $\omega$ (in rad/s) is related to the
wavenumber $\nu$ (in cm$^{-1}$) through $\nu = \omega/(200 \pi c )$ 
where $c$ is the speed of light (in m/s).  Modes are
orthonormalized using
\begin{equation}
\int \rho(\vec{R}) \vec{u}_{i}(\vec{R}) \cdot \vec{u}_{j}(\vec{R}) d^3\vec{R} = \delta_{ij} M_{p+m}
\label{orthonorm1}
\end{equation}
for any two modes $i$ and $j$, where $M_{p+m}$ is the mass of the system
(nanoparticle and matrix).  The integration is over the volume of the
system.  Equation (\ref{orthonorm1}) is justified in Appendix \ref{ApOrth}.
A general disturbance of the system has displacement field
\begin{equation}
\vec{u}(\vec{R},t) = \sum_n ( x_n sin(\omega_n t) + y_n cos(\omega_n t) ) \vec{u}_n(\vec{R}) 
 \label{aab}
\end{equation}
On the basis of the normalization chosen, $\vec{u}_n(\vec{R})$
is dimensionless. Therefore $x_n$ and $y_n$ have units of metres.

Inserting (\ref{aab}) into (\ref{kinetic}) and (\ref{potential})
and simplifying using (\ref{orthog1}) and (\ref{orthog2}),
the total energy of the system is
\begin{equation}
E = \frac12 M_{p+m} \sum_n \omega_n^2 (x_n^2 + y_n^2)
 \label{totalenergy}
\end{equation}

\section{Spherically Symmetric Models }

In this and later sections the displacement field 
depends on spherical material coordinates $\vec{R}$ = (R,$\theta$,$\phi$).
The nanoparticle ("$p$") is idealized as a homogeneous and
isotropic sphere of radius $R_p$ and 
density $\rho_p$ with speeds of sound $v_L^p$ and $v_T^p$.

In the free sphere model\cite{Lamb1882} (FSM),
the nanoparticle surface at $R = R_p$ is free.
Mode frequencies are $\omega^{FSM}_{lmnq}$ where
$q$ can be spheroidal (SPH) or torsional (TOR),
$l$ and $m$ are angular momentum and its z-component,
and $n$ is the harmonic index. For convenience,
these modes will be called "pseudomodes" 
when referring to situations of a sphere
that is only weakly coupled to its surroundings.
Several incorrect calculations of the FSM (SPH,$l$=$0$)
modes appear in the literature.\cite{saviotPRE04}

In the bound sphere model\cite{saviot96} (BSM),
the nanoparticle surface is rigidly fixed.
Mode frequencies are $\omega^{BSM}_{lmnq}$.

In the complex frequency model\cite{dubrovskiy81,saviotPRB03} (CFM),
the nanoparticle is surrounded by a homogeneous and isotropic
matrix ("$m$") of density $\rho_m$ and speeds of sound
$v_L^m$ and $v_T^m$.
The matrix extends to infinity.  Continuity of $\vec{u}$ and
force-balance apply at the nanoparticle-matrix interface.
The boundary condition at large $R$ is that $\vec{u}$
is an outgoing travelling wave.  The mode frequencies
$\omega^{CFM}_{lmnq}$ are complex valued.  In spectroscopic
observations of such a mode (a plot of scattered light intensity
versus wavenumber $\nu$) a broadened peak is observed
with its center at $\nu$ = Re($\omega$)/($200 \pi c$) and
with a half width at half maximum (HWHM) of Im($\omega$)/($200 \pi c$).

Other workers\cite{Tamura82,Ovsyuk96}
used the "absolute value of the spherical Hankel function" for
the outgoing wave, artificially reducing the mathematical problem
to real-valued frequencies.

There are two important limiting cases of CFM:  In the rigid
complex frequency method (RCFM) the nanoparticle is made
infinitely rigid and heavy.  In the void complex frequency
method (VCFM) the nanoparticle is replaced by an empty void.

CFM, RCFM and VCFM necessarily have complex-valued
$\vec{u}$ which blows up exponentially with $R$ and cannot be
orthonormalized.  Equations~(\ref{orthonorm1}) to (\ref{msd})
and the discussion in Appendix \ref{ApOrth} are for FSM, BSM and CSM
only and do not apply to CFM, RCFM and VCFM.

\begin{table}
  \begin{tabular}{|c|c|c|c|}
  \hline
  \backslashbox{matrix}{\\nano -\\particle} & none &  normal   & rigid heavy  \\
  \hline
   none        &      &    FSM    &      \\
  \hline
   normal      & VCFM & CFM , CSM & RCFM \\
  \hline
   rigid heavy &      &    BSM    &      \\
  \hline
  \end{tabular}

  \begin{tabular}{cc}
  FSM & Free Sphere Model\\
  BSM & Bound Sphere Model\\
  CFM & Complex Frequency Model\\
  RCFM & Rigid CFM\\
  VCFM & Void CFM\\
  CSM & Core-Shell Model
  \end{tabular}
\caption{\label{tab:abbr}Abbreviations used throughout this paper.}
\end{table}

\section{Core-Shell Model}
Adapting the core-shell model\cite{PortalesPRB02} (CSM)
we consider a nanoparticle as in the previous section concentrically
within a much larger homogeneous and isotropic sphere (representing
the glass matrix) with radius $R_m$.  In the macroscopic limit
$R_m$ goes to infinity, and the frequency of every vibrational
mode varies as $1/R_m$ and therefore goes to zero. Going from
the FSM to an embedded sphere with a macroscopic matrix means
changing from a discrete set of vibrational modes to a continuum.

We are primarily interested in Raman or Brillouin experiments
in which light is scattered
from the nanoparticle rather than from the transparent matrix
material. Low frequency vibrations of the nanoparticle allow
the scattered photon to have wavenumber a few cm$^{-1}$ above
or below those in the incident laser beam.  The general idea
behind this core-shell approach is that although the vibrations
of the whole system are described by a continuum of modes, the
nanoparticle doesn't move as much for matrix phonons of certain
wavelengths.

The inelastically scattered light is modulated by the
frequency-dependent amplitude of the vibration within the
nanoparticle. The mechanism for the photon scattering process
depends on the electronic details of the nanoparticle:  In
semiconductors, bound excitons; In metals, resonant plasmons;
In insulators, strain-optic coupling.  In any case, a
preliminary step is to study the amplitude of nanoparticle
vibrations as a function of phonon frequency. That is the
goal of our present investigation.

The first step is to solve the equation of motion for a given
value of $R_m$. Extending Takagahara's work\cite{takagahara96}
to core-shell systems, the normal modes have displacement
fields $\vec{u}_{l m n q}$ which are
normalized with respect to Eq.~(\ref{orthonorm1}).
Orthogonality of the modes was also checked numerically.

For the results reported here, the outer surface of the matrix
has zero traction force boundary conditions.
However, we have verified that use of zero displacement
boundary conditions has no noticeable consequence on the figures
presented in this paper.  Eigenfrequencies are affected, but
because we are interested in the macroscopic
limit ($R_m/R_p \rightarrow \infty$) this is not important.

To connect CSM with the FSM, BSM and CFM models, we compare
the discrete frequency sets to the position and width of peaks
of the mean square displacement inside the nanoparticle
(Eq.~(\ref{msd}) where $\vec{u}_{l m n q}$ is obtained from CSM)
as a function of frequency.

\begin{equation}
 \langle {u_{l m n q}}^2 \rangle_{p} = \frac1{\frac43 \pi R_p^3}
 \int_{R<R_p}
\| \vec{u}_{l m n q}(\vec{R}) \|^2
 d^3\vec{R}
 \label{msd}
\end{equation}

Using the selection rules predicted by Duval,\cite{Duval92} one can get
a rough idea of what the reduced Raman spectrum might look like:
only (SPH,$l$=0) and (SPH,$l$=2) modes are Raman active for a spherical
isotropic nanoparticle under non-resonant excitation.

The mean square displacement $\langle u^2 \rangle_p$
that we plot in our figures is only one of many possible measures
of the internal motion of the nanoparticle.  If we are interested
in low-frequency inelastic light scattering, then ultimately we
want to calculate the scattered photon spectrum.  Our model provides
a key ingredient for such a calculation in the $\vec{u}_{lmnq}(\vec{R})$.
Just as in plane wave vibrational modes, second quantization can
be applied to these modes to obtain phonon creation and
annihilation operators\cite{takagahara96} which would be required
in any model of photon scattering.

The $\langle u^2 \rangle_p$ variation with the mode frequency is
not affected once $R_m/R_p$ exceeds several times unity.
Increasing this ratio
results in more vibrational modes for a given frequency range. This
increase is compensated by the lower amplitude associated with each
mode. In this paper, we use $R_m/R_p=100$ when possible to have
continuous looking plots.
Calculations with bigger ratios were performed but don't change the
results. Note that $R_m/R_p=100$ means that
the nanoparticle occupies 1/1000000$^{th}$ of the total volume.

\section{Discussion}

\subsection{Ag in BaO-P$_2$O$_5$ glass}
Low-frequency inelastic light scattering\cite{portalesJCP01}
and femtosecond pump-probe experiments\cite{voisinPB02} on
nearly spherical Ag nanoparticles grown in a
BaO-P$_2$O$_5$ (high density glass) matrix show very well
resolved features, making it a perfect system to start with.

Elastic parameters for silver at 300K\cite{LBAg} are
directionally averaged\cite{saviotPRB03}
giving sound speeds 1740 m/s and 3750 m/s.
The actual transverse speed of sound in Ag varies
from 1195 m/s to 2080 m/s depending on the direction.
Longitudinal speed of sound varies from 3410 to 3935 m/s.
Rms averaged speeds (1760 m/s, 3750 m/s) are
comparable to the average speeds.
FSM vibrational frequencies calculated
using directionally averaged speeds
are in good agreement with frequencies calculated with
a molecular dynamic approach which fully takes into
account elastic anisotropy.\cite{saviotPRB03}
Mass density and sound speeds for the BaO-P$_2$O$_5$
matrix are taken from Voisin \textit{et al.}\cite{voisinPB02}

\begin{table}
 \begin{tabular}{|ccc|r|rr|rr|}
  \hline
       &   &   & FSM & \multicolumn{2}{c|}{CFM} & \multicolumn{2}{c|}{RCFM}\\
   $q$ & $l$ & $n$ & $\nu$  & $\nu$ & $\Delta\nu$ & $\nu$ & $\Delta\nu$\\
  \hline 
  TOR            & 2 & 0 &      &  2.6 & 5.8 &  2.4 & 4.2\\
                 &   & 1 &  4.7 &  7.2 & 1.0 &      &    \\
                 &   & 2 & 13.5 & 13.9 & 1.1 &      &    \\
                 &   & 3 & 19.8 & 20.1 & 1.1 &      &    \\
                 &   & 4 & 26.0 & 26.2 & 1.1 &      &    \\
  \hline
  TOR            & 1 & 0 &      &  0.0 & 4.5 &  0.0 & 2.8\\
                 &   & 1 &      &  4.9 & 1.1 &      &    \\
                 &   & 2 & 10.9 & 11.4 & 1.1 &      &    \\
                 &   & 3 & 17.1 & 17.5 & 1.1 &      &    \\
                 &   & 4 & 23.2 & 23.5 & 1.1 &      &    \\
                 &   & 5 & 29.2 & 29.4 & 1.1 &      &    \\
  \hline
  SPH            & 2 & 0 &      &  0.0 & 4.5 &  0.0 & 2.9\\
                 &   & 1 &      &  0.0 &12.0 &  0.0 &10.4\\
                 &   & 2 &  5.0 &  5.8 & 1.4 &      &    \\
                 &   & 3 &      &  8.8 & 9.2 &  8.2 & 7.5\\
                 &   & 4 &  9.7 & 11.1 & 1.2 &      &    \\
                 &   & 5 & 16.4 & 17.0 & 1.1 &      &    \\
                 &   & 6 & 21.9 & 22.4 & 1.6 &      &    \\
                 &   & 7 & 23.6 & 23.5 & 1.5 &      &    \\
                 &   & 8 & 29.0 & 29.2 & 1.1 &      &    \\
  \hline
  SPH            & 1 & 0 &      &  2.9 & 1.5 &      &    \\
                 &   & 1 &      &  4.5 & 8.0 &  4.1 & 6.4\\
                 &   & 2 &  6.9 &  8.4 & 1.2 &      &    \\
                 &   & 3 & 13.8 & 14.3 & 1.1 &      &    \\
                 &   & 4 & 17.3 & 17.6 & 1.9 &      &    \\
                 &   & 5 & 20.3 & 20.5 & 1.2 &      &    \\
                 &   & 6 & 26.2 & 26.4 & 1.1 &      &    \\
  \hline
  SPH            & 0 & 0 &      &  0.0 & 6.3 &  0.0 & 5.0\\
                 &   & 1 & 11.4 & 11.9 & 1.9 &      &    \\
                 &   & 2 & 24.9 & 25.1 & 1.9 &      &    \\
  \hline 
 \end{tabular}
 \caption{\label{tab:Ag}Wavenumber ($\nu$) and HWHM damping
 ($\Delta\nu$) in cm$^{-1}$ for free (FSM), \mbox{BaO-P$_2$O$_5$}
 embedded Ag nanoparticles (CFM) and rigid nanoparticle (RCFM) for modes
 with $\nu < $30~cm$^{-1}$ and $R_p$ = 4.9 nm.}
\end{table}

\begin{figure}[!ht]
 \includegraphics[width=\columnwidth]{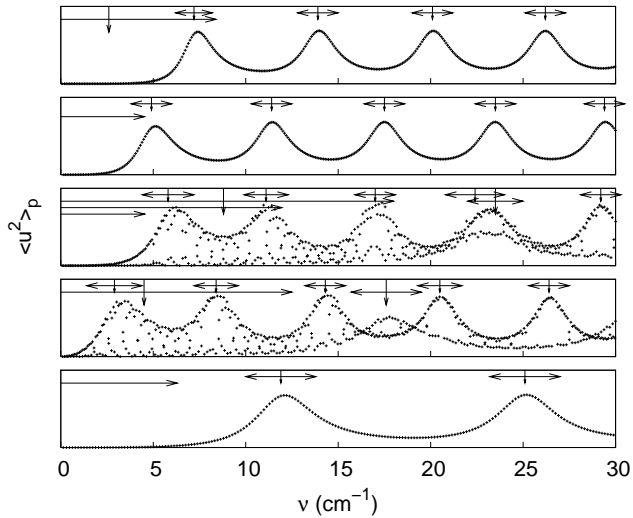}
 \caption{\label{u2Ag}Mean square displacement within the nanoparticle
 interior of vibrational modes for Ag $R_p$=4.9 nm in BaO-P$_2$O$_5$
 with $R_m/R_p=100$. SPH $l$=$0$,$1$ and $2$ and TOR $l$=$1$ and $2$ from
 bottom to top. Arrows indicate positions and FWHM obtained with the CFM.}
\end{figure}

Table~\ref{tab:Ag} shows FSM, CFM and RCFM wavenumbers
and HWHM damping.  As already reported for silicon
nanoparticles inside silica,\cite{saviotPRB03} there is good
agreement between CFM and FSM.
The main differences of CFM are: (1) nonzero damping and
(2) "extra modes" as described in the following section.
RCFM modes match some of the CFM modes not predicted
by FSM.

To present CSM results, the measure of nanoparticle
motion in Figs.~\ref{u2Ag} to \ref{u2CdS} is $\langle u^2 \rangle_p$,
the mean squared displacement within the nanoparticle of the
normalized vibrational mode displacement field as calculated
using Eq.~(\ref{msd}).  In each case,
$\langle u^2 \rangle_p$ is plotted individually for each
mode versus the mode's wavenumber in cm$^{-1}$.
To make it easier to compare different materials and different
sizes, the different $\langle u^2 \rangle_p$ plots cover
approximately the first two (SPH,$l$=0) pseudomodes.

Figure~\ref{u2Ag} shows individual plots for the separate
cases of $q$ and $l$.
Wavenumbers and HWHM of pseudomodes using CFM
as listed in Tab.~\ref{tab:Ag} are also indicated
on the figure as arrows.

It is apparent that the interior of the nanoparticle is 
in motion for every CSM mode, unlike FSM wherein the
nanoparticle would only vibrate at a discrete set of
frequencies.

Relatively smooth dependence
of $\langle u^2 \rangle_p$ on frequency
is seen for (TOR,$l$=1),
(TOR,$l$=2) and (SPH,$l$=0) modes.  The graph is much more complicated
for the (SPH,$l$=1) and (SPH,$l$=2) cases.  In these latter spheroidal
modes, the motion is a mixture of longitudinal and transverse
waves travelling at different speeds.  Normal modes are forced
to mix these two polarizations in order to satisfy
boundary conditions at $R = R_p$ and $R = R_m$.
Changing the boundary condition at $R = R_m$ from zero traction to zero
displacement doesn't change the appearance of these plots.

Points in the (SPH,$l$$\neq$$0)$ plots clearly
tend to cluster on two curves, with a few
points lying in between.
The (SPH,$l$$\neq$$0)$ plots consist of a longitudinal-like curve
with one broad peak and a transverse-like curve with five peaks.
CFM calculations also reveal this
feature because some modes are broader.

In agreement with Voisin \textit{et al.},\cite{voisinPB02}
Fig.~\ref{u2Ag} shows that Ag in BaO-P$_2$O$_5$ has
significant damping, but the nanoparticle's damped pseudomodes
are clearly visible.  This probably explains why well resolved
Raman peaks can be observed.\cite{portalesJCP01}

With the exception of "matrix modes" discussed in
the following section, there is good agreement between
the maxima of $\langle u^2 \rangle_p$ from CSM
and the wavenumbers and HWHM as predicted
using CFM. Where CFM predicts modes where
no $\langle u^2 \rangle_p$ peaks appear, these also
correspond to RCFM modes as discussed below. This
and subsequent figures demonstrate the utility of CFM
to predict the dominant frequencies at which the
nanoparticle oscillates as well as their width.

\subsection{Extra Modes}
The presence of the matrix leads to the appearance
of some qualitatively new modes not given by FSM.
We classify these as: (1) semi-rigid librational modes,
which appear at the low frequency end of the (TOR,$l$=1)
plots (near 5~cm$^{-1}$ in Fig.~\ref{u2Ag}).
Both CFM and CSM confirm their presence.  The
zero frequency limit of FSM for (TOR,$l$=1) corresponds
to rigid rotation.  We therefore identify low frequency
(TOR,$l$=1) modes as nearly rigid rotational oscillation.
(2) semi-rigid rattling modes, also at the low
frequency end of the (SPH,$l$=1) plots (near 3~cm$^{-1}$ in Fig.~\ref{u2Ag}),
also confirmed by both CFM and CSM.  The zero frequency limit
of FSM for (SPH,$l$=1) corresponds to rigid rectilinear motion.
We therefore identify low frequency (SPH,$l$=1) modes
as nearly rigid translational oscillation.
(3) "matrix modes" predicted by CFM but not apparent in
the plots of $\langle u^2 \rangle_p$ from CSM.  These
modes are well approximated by RCFM, which confirms their
nature as motions of the matrix which do not significantly
shake the nanoparticle.  VCFM is a good approximation for
the case of a heavier and more rigid matrix.

CFM roots were obtained numerically in this work.
However, some exact calculations are possible for "matrix modes".
As an example, for (SPH,$l$=0) modes,
there is only one VCFM eigenfrequency given by
$\xi = 2 \gamma\left(\sqrt{1-\gamma^2} + i \gamma\right)$
(see Eq.~(14'') in Dubrovskiy \textit{et al.}\cite{dubrovskiy81})
and only one RCFM eigenfrequency given by $\xi = i$
where $\gamma = {v_T^m}/{v_L^m}$ and
$\xi = {\omega R}/{v_L^m}$.
To our knowledge, this is the first report of Re($\omega$)=0 modes
in such systems.

\subsection{Pseudomode Damping}

In the absence of the matrix, the nanoparticle will vibrate
at frequencies $\omega^{FSM}_{lmnq}$. Coupling of the nanoparticle
to the matrix damps these pseudomodes.

One way to get a qualitative feeling for this problem is to
compare acoustic impedances.  For plane sound waves in a
fluid, acoustic impedance is density times speed of sound.
Reflection at normal incidence from a planar
interface is related to the difference of acoustic impedance
between the two fluids.  This concept also applies in general
terms to acoustic waves in a solid
(see Appendix \ref{impedance}).  When acoustic impedances
for the nanoparticle $Z_p=\rho_p v_L^p$ and the
matrix $Z_m=\rho_m v_L^m$ match, 
reflection of longitudinal sound waves is negligible.

\begin{figure}[!ht]
 \includegraphics[width=\columnwidth]{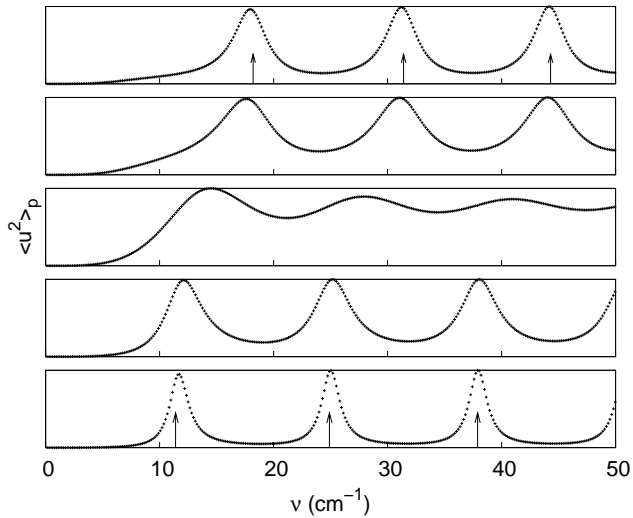}
 \caption{\label{rho}Mean square displacement
 within the nanoparticle interior
 of (SPH,$l$=0) CSM modes for a 4.9~nm
 radius silver nanoparticle and $R_m/R_p=100$. Matrix sound
 speeds are for BaO-P$_2$O$_5$. Matrix mass density is scaled
 from BaO-P$_2$O$_5$ by factors of 0.5, 1, 2.24, 4 and 6 from
 bottom to top. Arrows indicate the vibrational wavenumbers
 for the same silver particle with rigid
 (BSM, top) and free (FSM, bottom) boundary conditions. }
\end{figure}

As the acoustic impedance mismatch increases,
acoustic wave reflection increases leading to more pronounced
nanoparticle modes.  This is clearly seen in Fig.~\ref{rho} where
varying only the density of the matrix changes the locations
of frequencies at which $\langle u^2 \rangle_p$ peaks. For low
matrix mass density ($Z_p > Z_m$), $\langle u^2 \rangle_p$ peaks
close to the FSM frequencies. For high matrix mass
density ($Z_p < Z_m$), peaks appear close to the BSM
frequencies. For matching impedances
($\rho_m \approx 2.24\rho_{\text{BaO-P}_2\text{O}_5}$),
oscillations are weak.  

\subsection{Na in BaO-P$_2$O$_5$ glass}

This hypothetical case illustrates a soft nanoparticle
in a hard matrix.  Sodium is a very light and soft metal.
BaO-P$_2$O$_5$ is a heavy and rigid kind of glass.

Elastic parameters for Na at 300~K are from
P. Ho \textit{et al.}\cite{Ho68}
and directionally averaged to get sound speeds
1610~m/s and 3300~m/s.
In this case $Z_p/Z_m \approx 0.182$.

\begin{figure}[!ht]
 \includegraphics[width=\columnwidth]{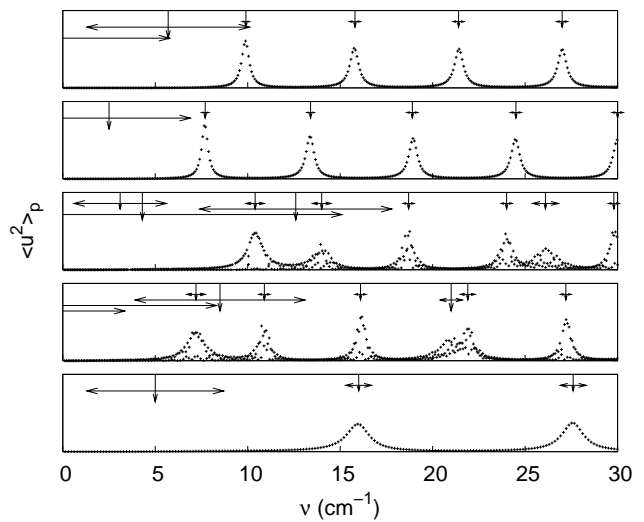}
 \caption{\label{u2Na}Mean square displacement
 within the nanoparticle interior of CSM vibrational modes
 for Na in BaO-P$_2$O$_5$ with $R_m/R_p=100$ and $R_p=4.9$ nm.
 SPH $l$=$0$,$1$ and $2$ and TOR $l$=$1$ and $2$
 from bottom to top. Arrows indicate positions and FWHM obtained 
 from CFM.}
\end{figure}

Figure~\ref{u2Na} shows the plots of $\langle u^2 \rangle_p$ from CSM.
CFM mode wavenumbers and HWHM are shown as arrows.
Peaks of $\langle u^2 \rangle_p$ are always narrow and
close to BSM frequencies (not shown). CFM predicts additional
much broader peaks which do not appear in $\langle u^2 \rangle_p$,
but these are close to VCFM frequencies as given in Tab.~\ref{tab:matrix}.
Unlike a heavy nanoparticle in a light matrix, no additional
low frequency mode corresponding to semi-rigid
librational (TOR,1) or rattling (SPH,1) mode appears in the
$\langle u^2 \rangle_p$ plots.

\begin{table}
  \begin{tabular}{|ccc|dd|dd|dd|}
  \hline
       &     &     &\multicolumn{2}{c|}{VCFM}\\
  mode & $l$ & $n$ & \multicolumn{1}{c}{$\nu$}
		   & \multicolumn{1}{c|}{$\Delta\nu$}\\
  \hline
  SPH     & 0 & 0 &  4.6 &  3.2 \\
          & 1 & 0 &  0.0 &  3.1 \\
          &   & 1 &  0.0 &  7.9 \\
          &   & 2 &  7.9 &  4.4 \\
          & 2 & 0 &  2.9 &  2.3 \\
          &   & 1 &  4.1 & 10.4 \\
          &   & 2 & 11.9 &  5.0 \\
  \hline
  TOR     & 1 & 0 &  2.4 &  4.2 \\
          & 2 & 0 &  0.0 &  5.6 \\
          &   & 1 &  5.4 &  4.2 \\
  \hline
  \end{tabular}
\caption{\label{tab:matrix}Wavenumber ($\nu$) and damping
($\Delta\nu$) in cm$^{-1}$ of modes of a BaO-P$_2$O$_5$ matrix
surrounding a spherical cavity (VCFM) with radius $R_p=4.9$ nm.}
\end{table}

\begin{table}
  \begin{tabular}{|ccc|dd|dd|dd|}
  \hline
       &     &     &\multicolumn{2}{c|}{SiO$_2$}
                   & \multicolumn{2}{c|}{GeO$_2$}\\
  mode & $l$ & $n$ & \multicolumn{1}{c}{$\nu$}
                   & \multicolumn{1}{c|}{$\Delta\nu$}
		   & \multicolumn{1}{c}{$\nu$}
		   & \multicolumn{1}{c|}{$\Delta\nu$}\\
  \hline
  SPH     & 0 & 0 &  0.0 &  9.3 &  0.0 &  5.4\\
          & 1 & 0 &  7.6 & 12.2 &  4.4 &  7.0\\
          & 2 & 0 &  0.0 &  6.2 &  0.0 &  3.5\\
          &   & 1 &  0.0 & 19.6 &  0.0 & 11.3\\
          &   & 2 & 15.2 & 14.5 &  8.8 &  8.3\\
  \hline
  TOR     & 1 &   &  0.0 &  5.9 &  0.0 &  3.3\\
          & 2 & 0 &  5.1 &  8.8 &  2.9 &  5.0\\
  \hline
  \end{tabular}
\caption{\label{tab:softmatrix}Wavenumber ($\nu$) and FWHM ($\Delta\nu$)
in cm$^{-1}$ of a matrix surrounding an infinitely rigid and heavy sphere
(RCFM). Sphere radius is 3.4 nm.  }
\end{table}

\subsection{Si in SiO$_2$}

This example is chosen to show what happens when the elastic
properties of the nanoparticle and the matrix are very similar.
Low frequency Raman experiments on such
systems\cite{Fujii96,pautheJMC99,saviotPRB03}
reveal broad peaks whose Raman position scales inversely with
nanoparticle size.

\begin{figure}[!ht]
 \includegraphics[width=\columnwidth]{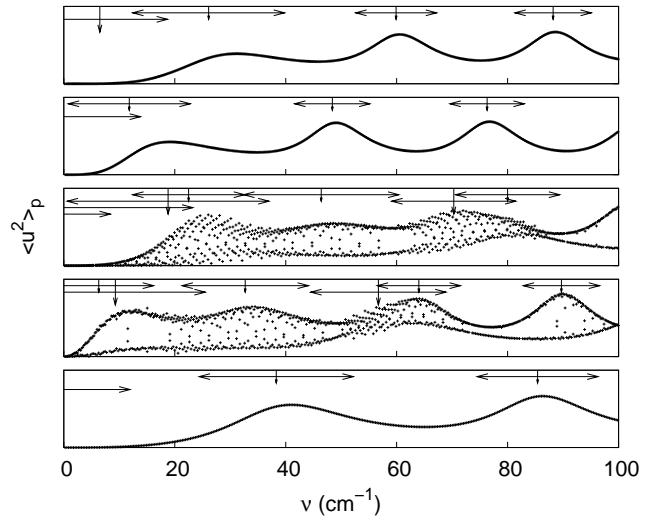}
 \caption{\label{u2Si}Mean square displacement
 within the nanoparticle interior of CSM vibrational
 modes for Si in SiO$_2$ with $R_m/R_p=100$ and $R_p=3.4$ nm.
 SPH $l$=$0$,$1$ and $2$ and TOR $l$=$1$ and $2$ from bottom
 to top. Arrows indicate positions and FWHM obtained from CFM.}
\end{figure}

Elastic parameters and mass density for silicon and silica
are taken from our previous work.\cite{saviotPRB03} $R_p=3.4$ nm
comes from the same paper and we chose to use the same value for
the following systems.  In contrast to Ag in BaO-P$_2$O$_5$
where $Z_p/Z_m \approx 2.34$, in the case of silicon inside silica, we
have $Z_p/Z_m \approx 1.60$.
As can be seen in Fig.~\ref{u2Si}, this results in blurred features.
Silicon has a lower sound speed ratio ($v_L/v_T \approx 1.67$)
than silver ($2.15$). Therefore fewer TOR and (SPH,$l$$\neq$$0$)
pseudomodes are seen in the figure. In a previous
work,\cite{saviotPRB03} we attributed the lowest observed
low-frequency inelastic light scattering peak to scattering by
the first (TOR,$l$=1) pseudomode. At the same time, we pointed
out that the calculated damping was too high. From the (TOR,$l$=1)
plot, we can see that a proper description of the scattering
process is necessary to model the Raman spectrum. In particular,
the role of the electronic levels need to be taken into account.
At the same time, a better description of the silicon-silica
interface could somewhat mechanically isolate the silicon
nanoparticle which would result in narrower pseudomodes.
Introducing a thin intermediate layer between the nanoparticle
and the matrix is a way to achieve this.\cite{PortalesPRB02}
However, choosing the mass density and elastic parameters for
this intermediate shell is a difficult task.

\subsection{CdS} % Both CdS in the same section now

Many low-frequency Raman studies have been published for
matrix embedded CdS$_x$Se$_{1-x}$ nanoparticles. Because of the
different matrices used and varying nanoparticle
composition, it is very difficult to address all of them here.
In this paper, we will focus on CdS nanoparticles and show the
influence of two different matrices. Low-frequency Raman
spectra of CdS nanoparticles embedded in sol-gel silica have been
reported by Othmani \textit{et al.}\cite{Othmani93} and
Saviot \textit{et al.}\cite{saviotPRB98}
GeO$_2$ embedded CdS nanoparticles were investigated by
Tanaka \textit{et al.}\cite{Tanaka93}

Elastic parameters for CdS at 300~K are taken from
Berlincourt \textit{et al.}\cite{Berlincourt63}
and directionally-averaged as before to get
sound speeds 1870~m/s and 4290~m/s.
We use elastic constants measured at constant
electric field and do not consider piezoelectric
corrections.  Mass density and sound speeds for
amorphous GeO$_2$ are from Antoniou \textit{et al.}\cite{Antoniou65}

\begin{figure}[!ht]
 \includegraphics[width=0.7\columnwidth]{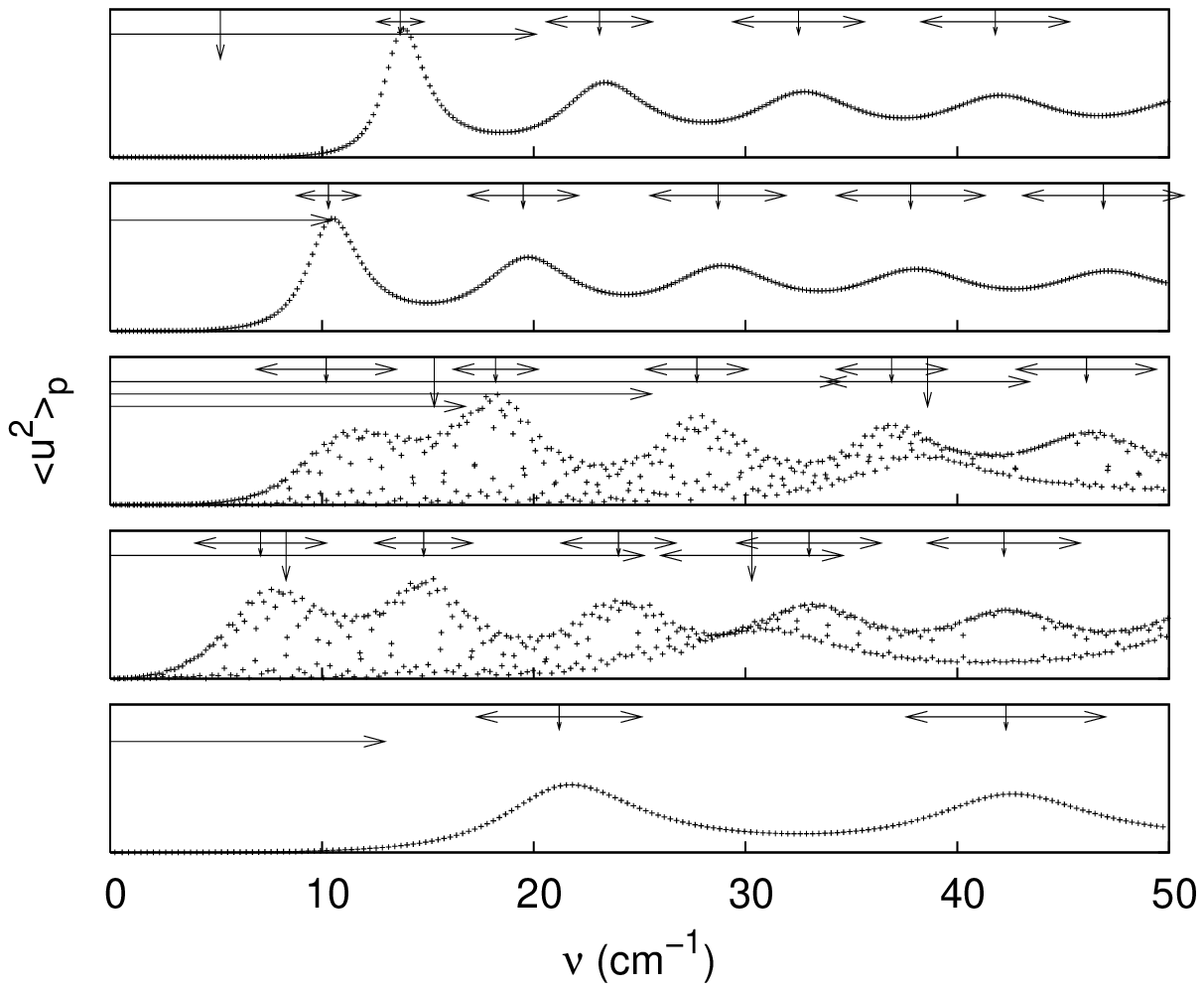}
 \includegraphics[width=0.7\columnwidth]{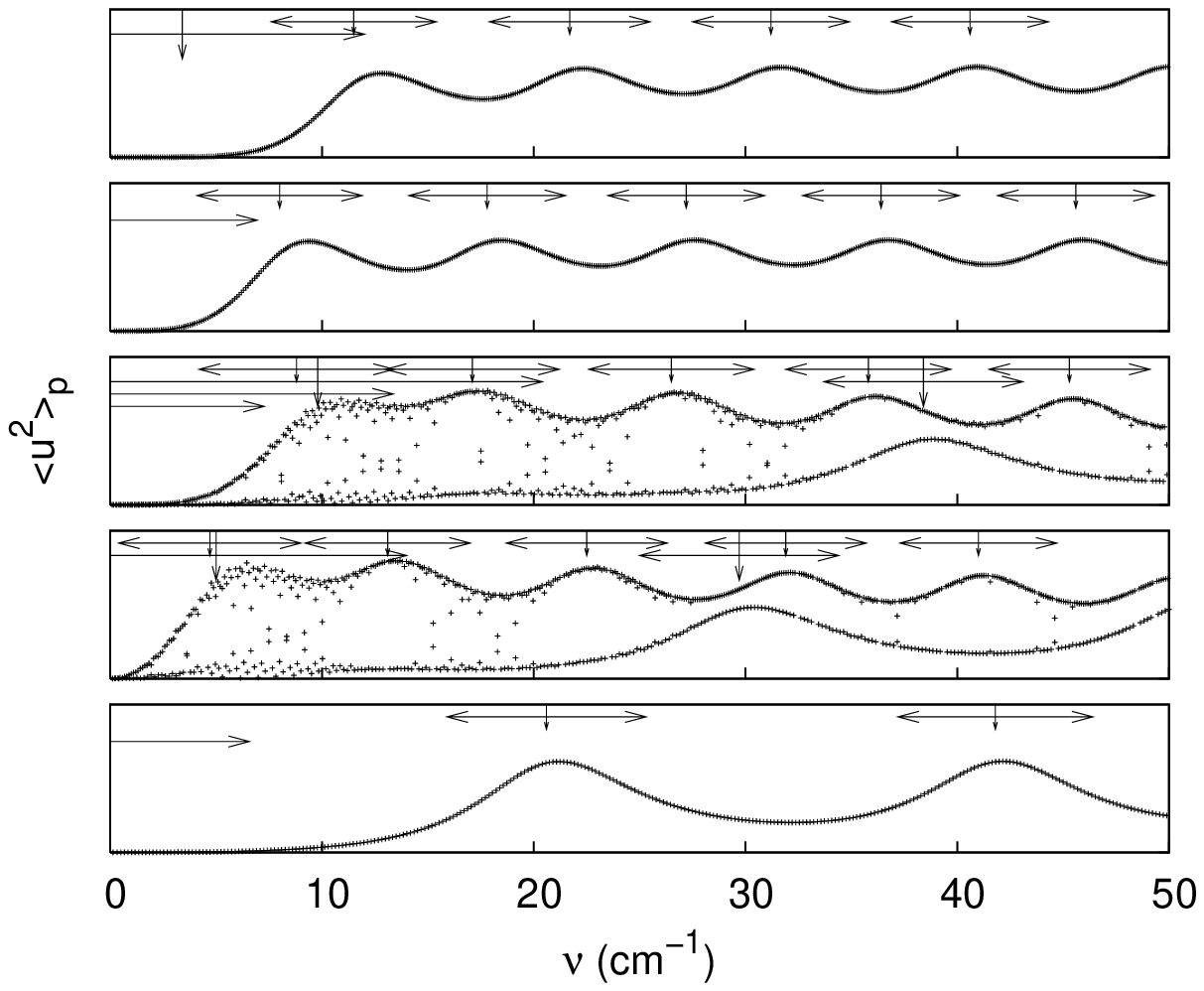}
 \caption{\label{u2CdS}Mean square displacement
 within the nanoparticle interior of vibrational modes for CdS in
 SiO$_2$ (top) and GeO$_2$ (bottom) with $R_m/R_p=100$ and $R_p=3.4$ nm.
 SPH $l$=$0$,$1$ and $2$ and TOR $l$=$1$ and $2$ from bottom to top. Arrows
 indicate CFM positions and FWHM.}
\end{figure}

CSM $\langle u^2 \rangle_p$ plots and CFM results
are presented in Fig.~\ref{u2CdS}.
In both cases $Z_p/Z_m \approx 1.6$
which is very close to the previous Si in SiO$_2$ system.
GeO$_2$ having a smaller impedance than SiO$_2$ (even when using
transverse sound speeds), one would expect the features of the GeO$_2$
embedded nanoparticles to be slightly more pronounced than the SiO$_2$ ones
because the impedance mismatch is bigger. It turns out to be more complex.
Modes having a strong longitudinal character follow this trend while those
having a strong transverse character present more resolved low-frequency
features for SiO$_2$. This is revealed by both the CFM and CSM calculations.
Notably, in the SiO$_2$ matrix, both CSM and CFM show a
sharp (TOR,$l$=1) rigid librational mode at 10.3~cm$^{-1}$.
The hexagonal crystal structure
and lack of inversion symmetry of CdS makes its electronic
properties anisotropic even if the nanocrystal is a perfect sphere.
So it is possible that this mode could be Raman active.
Acoustic properties are anisotropic too, even if this
paper works around this issue by performing a directional average.
Rigid librations may therefore have the ability to
couple to electronic degrees of freedom and modulate
inelastic light scattering.

The only satisfactory way to explain the low-frequency Raman features
is to couple the CSM approach with the work done by Sirenko
\textit{et al.}\cite{Sirenko98} on the interaction of the electronic
excitations quantized in the confining potential of the nanoparticle
with bulk-like acoustic phonons. The calculated Raman spectra
should be further narrowed by taking into account the variation of the
vibration amplitude inside the nanoparticle as a function of the
vibration frequency.

\subsection{Vibrational density of states}

Free\cite{Pasquini03} and weakly coupled\cite{Geller03} nanoparticles are
interesting objects because their vibrational density of states (VDOS)
below the lowest FSM mode vanish. However, in our case both FSM and
CFM have discrete mode spectra and do not reproduce the correct phonon
density of states.  Using CSM we calculated the VDOS. Note that this
only involves finding the frequencies of the core-shell system. Amplitude
normalization is not needed here.

\begin{equation}
 \frac{N(\nu)}{V}
   =
   \frac{1}{V} \int_0^\nu D(\nu^\prime) d\nu^\prime \\
 \label{Nw}
\end{equation}
where $V$ is the volume of the core-shell system.

\begin{figure}[!ht]
 \includegraphics[width=\columnwidth]{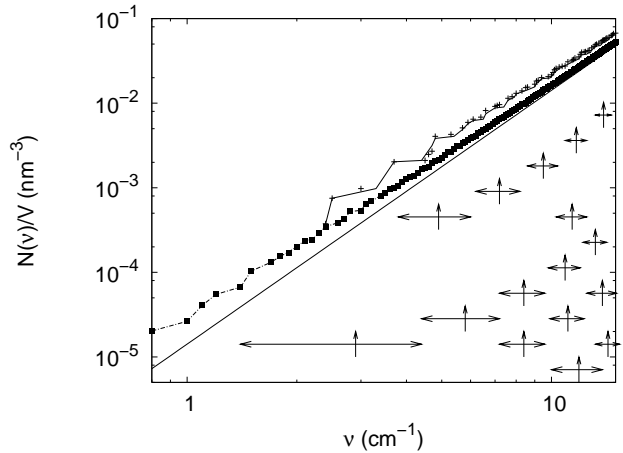}
 \caption{Cumulative number of phonon modes per unit volume
 as a function of wavenumber for a
 silver nanoparticle inside BaO-P$_2$O$_5$ with $R_m/R_p=3$ (plus symbols)
 and $R_m/R_p=10$ (square symbols). The corresponding matrix only points
 are connected with lines and are very close to the previous symbols.
 The straight line corresponds to the Debye model (Eq.~(\ref{debye})).
 Position and width of the nanoparticle CFM pseudomodes are
 indicated by arrows: SPH $l$=$0$,$1$,$2$,$3$,$4$,$5$ and
 TOR $l$=$1$,$2$,$3$,$4$,$5$ from bottom to top respectively.
 "Matrix modes" are not shown.}
 \label{N}
\end{figure}

The cumulative number of modes per unit volume up to wavenumber $\nu$
(Eq.~(\ref{Nw})) is plotted in Fig.~\ref{N} for the case of a
silver nanoparticle embedded in a BaO-P$_2$O$_5$ matrix.
The lowest frequency for a given mode ($q$,$l$) increases
with $l$. The numbers of TOR and SPH modes up to $l$=$60$
had to be summed taking into account the $2l+1$ degeneracy.
In this way, the VDOS up to $\nu=15$~cm$^{-1}$ for $R_m/R_p=10$
was obtained. Calculations for $R_m/R_p=3$ were also performed.
Results are compared to the Debye model using matrix sound speeds
(Eq.~(\ref{debye})).
\begin{equation}
 \frac{N_{Debye}(\omega)}{V}
   =
   \frac{1}{2\pi^2} \left( \frac{1}{v_L^3}+\frac{2}{v_T^3} \right)
     \frac{\omega^3}{3}
 \label{debye}
\end{equation}

The presence of the nanoparticle doesn't appreciably
modify the VDOS, even for small $R_m$.

\section{Conclusion}

Within a continuum elastic approximation, the phonon spectrum
of an isotropic spherical nanoparticle embedded in a macroscopic
matrix can be calculated without approximation.  This facilitates
computations that require information about the phonon modes,
and resolves the apparent contradiction involved in using free
sphere vibrational modes to describe the mechanical oscillations
of embedded nanoparticles.

Even though by necessity an object embedded in a macroscopic
matrix must vibrate at every frequency, the vibration of a
nanoparticle in a light and soft matrix is clearly dominated by
FSM modes. This explains the success of FSM to reproduce the
results of Raman scattering experiments on metal and semiconductor
nanoparticles embedded in glasses.

CFM is demonstrated to accurately predict both the frequency and
HWHM of pseudomodes of a matrix-embedded nanoparticle.  New low
frequency features corresponding to semi-rigid librational and
rattling motions as a result of the restorative force of the
matrix are clearly confirmed using both CFM and CSM.  This points
out the most serious limitation of FSM for embedded nanoparticles.
Even so, FSM-predicted frequencies of pseudomodes turn out to be
surprisingly accurate.
It is the acoustic impedance mismatch between nanoparticle and
matrix due both to density and speed of sound difference that
allows dampening to be sufficiently slight
that nanoparticles can "look free" in some light scattering
experiments.

While the new (SPH,$l$=1) and (TOR,$l$=1) modes are not Raman
active for an isotropic perfectly spherical nanoparticle, various
imperfections can potentially allow these motions to modulate the
polarizability of the nanoparticle and contribute to Raman and
Brillouin spectra.  However, in situations where elaboration
methods allow relatively spherical nanoparticle growth, these new
modes remain "invisible" in Raman and Brillouin experiments.
"Matrix modes" are also present, but because they are mainly
restricted to the matrix, they can't be observed in experiments
which are very often made under resonant conditions on the
nanoparticle electronic transitions.  Thus, the new qualitative
features of the motion of embedded nanoparticles are not readily
observable in experiments.

Size variation within a sample of nanoparticles causes a
distribution of nanoparticle pseudomode frequencies.  Sample
characterization methods such as TEM have the ability to measure
the extent of this size variation, but elaboration methods are
normally limited in their ability to restrict the range of
nanoparticle radius within a sample.  This important effect
"masks" whatever line broadening of the pseudomodes results from
matrix embedding.

Taken together, every feature of these systems conspires
to obscure experimental indications that the nanoparticles
really are embedded in solid material as opposed to
floating in a vacuum.  Consequently the original 1882 Lamb
FSM remains a powerful calculational tool
for the description of nanoparticle vibrations.

Displacement fields using CFM are complex valued and also blow
up exponentially with $R$.  Also, the frequencies are complex
valued.  So these cannot be used as classical degrees of freedom
to be quantized.  CSM provides a rigorous foundation for more
detailed theoretical studies since it is clear how it can be
quantized.  Work is in progress to apply this approach together
with the proper description of the nanoparticle confined
electronic levels in order to simulate the Raman spectrum of
some systems.

\appendix

\section{Mode Orthogonality}
\label{ApOrth}
The kinetic energy and potential energy of the system (assuming only mode A is excited) are:
\begin{equation}
T_A(t) = \int \frac12 \rho(\vec{R})
\| \dot{\vec{u}}_A (\vec{R},t) \|^2
d^3\vec{R}
\label{kinetic}
\end{equation}
\begin{equation}
V_A(t) = \int \frac12 c_{ijkl} u_{Ai,j} u_{Ak,l} d^3\vec{R}
\label{potential}
\end{equation}

By conservation of energy $E_A$ = $T_A(t)$ + $V_A(t)$
cannot depend on time.
Suppose now that two modes A and B ($\omega_A \neq \omega_B$)
are simultaneously excited, so that

\begin{equation}
\vec{u}(\vec{R},t) = sin(\omega_A t) \vec{u}_A(\vec{R}) + sin(\omega_B t) \vec{u}_B(\vec{R})
\end{equation}

$T_{A+B}(t)$ is the kinetic energy due to the superposition
of the two modes.  Therefore,

\begin{eqnarray}
T_{A+B}(t)  & = & T_A(t) + T_B(t) \\
         & + & \omega_A \omega_B cos(\omega_A t) cos(\omega_B t) \int \rho 
      \vec{u}_A \cdot \vec{u}_B d^3\vec{R} \nonumber
\end{eqnarray}

$V_{A+B}(t)$ is the elastic potential energy due to the
superposition of the two modes.  So,

\begin{eqnarray}
V_{A+B}(t) & = & V_A(t) + V_B(t) \\
           & + & sin(\omega_A t) sin(\omega_B t)
\int c_{ijkl} u_{Ai,j} u_{Bk,l} d^3\vec{R} \nonumber
\end{eqnarray}

With both modes excited, the energy is $E_{A+B}$.
We next evaluate $E_{A+B} - E_A - E_B$.
By energy conservation
(since there are no body or traction forces or energy dissipation mechanisms)
$E_A$, $E_B$ and $E_{A+B}$ must all be constant in time.
Consequently,

\begin{equation}
\int \rho(\vec R) \vec{u}_A \cdot \vec{u}_B d^3\vec{R} = 0
 \label{orthog1}
\end{equation}

\begin{equation}
\int c_{ijkl} u_{Ai,j} u_{Bk,l} d^3\vec{R} = 0
 \label{orthog2}
\end{equation}

Equation~(\ref{orthog1}) was derived as equation (3.72)
on page 39 of Trallero-Giner \textit{et al.}\cite{polarmodes}

Each of Eq.~(\ref{orthog1}) and Eq.~(\ref{orthog2}) provides
an orthogonality on
nondegenerate normal modes of vibrations.  For convenience,
we choose to orthonormalize the normal modes as
in Eq.~(\ref{orthonorm1}).  Within a degenerate subspace,
Gram-Schmidt orthogonalization can be used with this inner
product to generate a complete orthonormal basis of normal
modes.

\section{Energy Equipartition}
\label{pseudo}

Nanoparticle vibrations are very far from thermal equilibrium
when a short pulse of laser light heats the nanoparticle to
thousands of degrees.  Conversely, many experiments are done
with low incident laser beam intensities so that temperature
is constant throughout the nanoparticle and surrounding matrix.

For a linear classical mechanical system in thermodynamic
equilibrium, each degree of freedom has average energy $k_B T$.
Here we consider the approximate applicability of this
equipartition theorem to vibrations of the embedded nanoparticle.

The nanoparticle is not an isolated system because of its mechanical
connection to the matrix, but if the density or speeds of sound
are significantly different (between nanoparticle and matrix)
then there may be in effect a poor
impedance match at the interface so that vibrational energy
cannot readily enter or exit the nanoparticle.  In this case
there is hope that the nanoparticle can be
approximated as a nearly isolated mechanical system only weakly
coupled to the matrix.

Removing the matrix, the normal vibrational "pseudomodes" of
the nanoparticle can be indexed by $l m \tilde{n} q$.  The term
"pseudomodes" emphasizes that they are not normal modes because
of the non-isolation of the system.  Three indices $l$, $m$ and
$q$ are in common with the corresponding indices of modes of the
core-shell system as a whole, $\tilde{n}$ is the index for
the nanoparticle pseudomodes and $n$ is the index for
normal modes of the entire core-shell system.

Assuming weak coupling, each pseudomode $l m \tilde{n} q$ will
have average kinetic energy $\frac12 k_B T$.
The coupling to the matrix leads to thermal fluctuations
both of amplitude and phase of the motion
within the nanoparticle.  As a result the motion of a single
pseudomode has a widened band of frequencies.

We add up kinetic energies within
the nanoparticle for all core-shell modes 
whose $\omega_{l m n q}$ is close to 
$\omega_{l m \tilde{n} q}$ of the pseudomode.  This gives
a time average kinetic energy associated with the pseudomode:

\begin{equation}
\langle T_{l m \tilde{n} q} \rangle_{t} =
\sum_{n} \frac{1}{4} \omega^2_{lmnq} ( x^2_{lmnq} + y^2_{lmnq} )
M_{p} \langle u^2_{lmnq} \rangle_{p}
\end{equation}

Since each core-shell mode has average energy $k_B T$,
$x^2_{lmnq} + y^2_{lmnq}$ equals
$ 2 k_B T/( M_{p+m} \omega_{lmnq}^2 )$.
Thus,
\begin{equation}
\sum_n \langle u^{2}_{lmnq} \rangle_{p}
\approx
\frac{M_{p+m}}{M_{p}}
\end{equation}
where the summation over $n$ includes only such core-shell
modes for which $\omega_{lmnq}$ is close to a particular
pseudomode frequency $\omega_{l m \tilde{n} q}$.
This is well satisfied for the cases we have checked.

Since normal modes are normalized with respect to
Eq.~(\ref{orthonorm1})
and since the integrand of (\ref{msd}) is non-negative,
$\langle u_{lmnq}^2 \rangle_{p} \leq {M_{p+m}}/{M_{p}}$.

\section{Acoustic Impedance}
\label{impedance}

Acoustic impedance for sound waves can be generalized to
the situation of a spherical wave crossing a spherical
surface at normal incidence.  In this case the acoustic
impedance is complex valued.  In
addition to density and speed of sound it also depends
on frequency and the radius of the interface.  When the
interface radius is large compared to the wavelength the
acoustic impedance again approaches density times speed
of sound.  See equation (5.11.12) on page 128 of
L. E. Kinsler \textit{et al.}\cite{Kinsler}

\bibliography{coreshell}

\end{document}